# Recognition of paper samples by correlation of their speckle patterns

Complex Photonic Systems, University of Twente
P.O. Box 217
7500 AE Enschede, The Netherlands
E-mail: a.p.mosk@utwente.nl

October 12, 2006

Frerik van Beijnum
Elbert van Putten

Under the supervision of:
Karen van der Molen
Allard Mosk

## Abstract

Buchanan *et al.* [Nature **436,** p. 475 (2005)] have shown that it is possible to recognize paper samples via their speckle pattern by using a line-shaped laser focus, four photo detectors and a scanning mechanism. In this report recognition of five out of ten paper samples is presented. The sample was illuminated by a 2.62 ± 0.02 mm circular spot and the reflected light was measured by a static CCD.

    We have formulated a criterion for recognition that limits the probability of false recognition to 0.1 % for the experiment with ten samples. We obtained results that show that the probability of false recognition will be negligible for a large amount of samples.

    The properties of speckle originating from a line and a spot illumination source have been compared to see whether the use of a line results in major advantages for the recognition of samples. For this comparison a line (69 ± 1 μm width and 1.93 ± 0.01 mm length) and a spot (365 ± 5 μm diameter) were used. In all correlations this line shows a stronger correlation peak than the spot when the size of the peak is measured relative to the standard deviation of the correlation.

# Preface

The experiment described in this report was performed during a course at the Complex Photonic Systems group of Willem Vos and Ad Lagendijk at the University of Twente. To gain experience in performing research the group gave us the opportunity to work in their laboratory. Inspired by an article of Buchanan *et al.* we have chosen the subject 'recognizing samples via their speckle pattern'. We experienced the given opportunities in the lab and the freedom of choosing our own assignment as very challenging and motivating, for this we would like to thank the coordinator of the course, Allard Mosk.

   In our every day work we enjoyed being part of the group. We have attended the group meetings which gave us an impression of the research performed in the group. These meetings were also used to present our results; which taught us how to present results in a better way.

   Within the group our two fellow students Ramy El-Dardiry and Sanli Faez were of great importance for the atmosphere in which we worked. Both students are very motivated in their work and they had many good ideas, this inspired us in our project.

   Karen van der Molen and Cock Harteveld have guided us on a daily basis. Cock was always available for practical or technical questions. Besides this, he often took a glance at our setup and gave us many good suggestions on improving it. Karen has always been interested in the progress of our work. We experienced her critical attitude as very motivating. Furthermore she showed us how everyday scientific work goes. In the ups and downs of the research her support was very welcome.

Thank you all, it was a great pleasure,

  Elbert and Frerik

# Contents

# 1 Introduction

In our world forgery is timeless; ever since money and important documents exist falsifications are made. Many efforts have been done to distinguish these falsifications from real items. Buchanan *et al.* [1] have shown that paper samples can be distinguished by their unique internal structure. This structure can be visualized by illuminating the paper sample, this yields a unique speckle pattern which can be measured and stored. When these measurements are compared to other measurements, samples can be distinguished. Reproducibility of the speckle pattern is an important issue when using the pattern to distinguish samples. A speckle pattern is reproducible when the random structure of the paper does not change in time. Furthermore it is important that the speckle pattern can be measured in a reproducible way.

Buchanan *et al.* have distinguished 500 paper samples by their speckle pattern. The recognition is even possible after several ways of rough handling, such as baking it in an oven, soaking and heavily screwing the sample. Buchanan *et al.* measured the speckle pattern of the reflected light with a scanning mechanism, a laser line and four photo detectors. The experiments described in this report omit the scanning mechanism and the photodiodes. One static Charge Coupled Device (CCD) was used to measure the speckle pattern of the reflected light. A large circular spot was used to create a speckle pattern for the recognition attempt presented in this report. Measurements to find advantages of using a line instead of a spot were performed too.

In this report we first describe the theoretical aspects of our experiments: In chapter 2 an estimation of the speckle size, Gaussian beam theory, an analysis of the properties of speckle and the used recognition criterion are presented. A discussion of the used recognition criterion we concludes the chapter. In the third chapter the setup on which all measurements were performed will be presented. The chapter 'Results' will present all final results, and discusses the performed measurements. Plots on sample movements, distance variations and the angular sensitivity are presented. From these measurements we will conclude that the use of a line has no significant advantages for the recognition of samples. A presentation of the recognition of the samples concludes this chapter. In the last two chapters the conclusions and recommendations will be given.

# 2 Theory

This chapter deals with a few concepts that are used throughout the report. In section 2.1 the origin of speckles is discussed together with an estimation of the average size of speckles. Section 2.2 discusses Gaussian beam optics which is necessary for creating a spot and a line of a certain size. In section 2.3 the behavior of the speckles is discussed when a degree of freedom is changed. We conclude the chapter with a description of our recognition criterion.

## *2.1 Speckles*

When highly coherent light falls onto an object and the scattered light is projected onto a screen, the screen is speckled with bright and dark regions. These speckles are caused by rays scattering from different parts of the illuminated area. At the screen these rays have a different optical path length; therefore the rays interfere and result in speckles, as shown in Figure 2.1. In transmission speckles will be found too, in this case the phase shift between the rays arises from several scattering events within the sample.

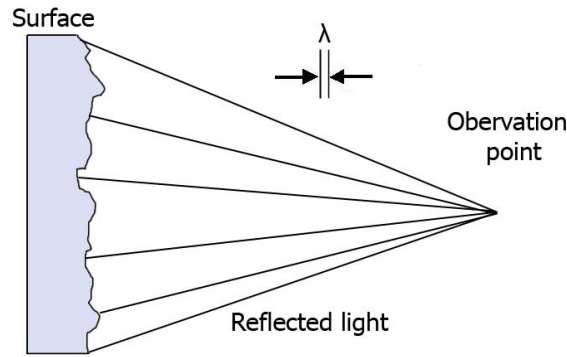

Figure 2.1: Rays reflect from different parts of the surface. The surface is optically rough thus the rays have a different optical path length at the observation point and interference will give bright and dark regions, so-called speckles.

The relation between the illuminated part of the sample and the average size $\sigma$ of the resulting speckles can be estimated by

$$\sigma \approx \frac{2\lambda L}{D}, \qquad (2.1)$$

where $\lambda$ is the wavelength of the laser beam, $L$ the distance from the sample to the screen and $D$ the length of the illuminated part of the sample. In Appendix A the derivation of this order-of-magnitude estimation for the speckle size is given. Equation (2.1) shows that when a larger area is illuminated the resulting speckles will shrink. The equation also shows that a sample lit by a horizontally aligned laser line will have a speckle pattern consisting of vertically aligned lines.

Increasing the power of the light source increases the spread in intensity of the speckles since the highest measured intensity increases while the minimum remains zero.

## *2.2 Gaussian beam optics*

The electromagnetic field modes inside most optical laser cavities can be described mathematically by TEM$_{mn}$ (Transverse Electric and Magnetic) modes, where m and n represent the order of the modes perpendicular to the propagation direction. The fundamental mode (*m=n=0*) has a Gaussian intensity profile. The HeNe laser that will be used is considered to be an ideal TEM₀₀ mode [3, 4]; therefore Gaussian beam optics will be discussed now.

Due to the fact that no real laser beam can be infinite in size, diffraction causes beams to spread transversely as they propagate. The radius of a Gaussian beam $w$, defined as the distance at which the intensity drops to e$^{-2}$ times the maximal axial value, spreads in accordance to

$$w(z) = w_0 \left[ 1 + \left( \frac{\lambda z}{\pi w_0^2} \right)^2 \right]^{1/2}, \qquad (2.2)$$

where $w_0$ is the minimal radius of the beam, the so-called beam waist, which is depicted in Figure 2.2. The wavelength is given by $\lambda$. The distance with respect to the beam waist is given by $z$.

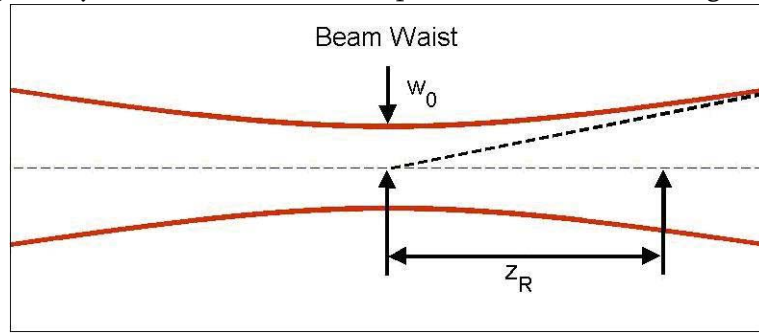

Figure 2.2: Growth in beam diameter as a function of distance from the beam waist.

A practical measure for the divergence is the distance $z_R$ with respect to the focus. At $z_R$ the cross-sectional area of the beam is doubled. Using equation (2.2) and setting $w(z)$ equal to $\sqrt{2} \cdot w_0$ one can solve this for $z_R$:

$$z_R = \frac{\pi w_0^2}{\lambda}. \qquad (2.3)$$

The distance $z_R$ is widely known as the Rayleigh range and $z_R$ can be considered as a measure for the distance at which the beam waist remains rather constant, and in focus.

## *2.3 Analyses of changes in the speckle pattern*

Recognizing a sample using the speckle pattern of two different measurements requires that the sample is mounted at the same place with certain accuracy. In this section an analysis is given of the changes in the speckle pattern caused by changing a degree of freedom in the setup. The sideward ($x$), displacement of the sample, the distance between sample and CCD ($r$) and the angular displacement ($\theta$) of the CCD are varied, see Figure 2.3.

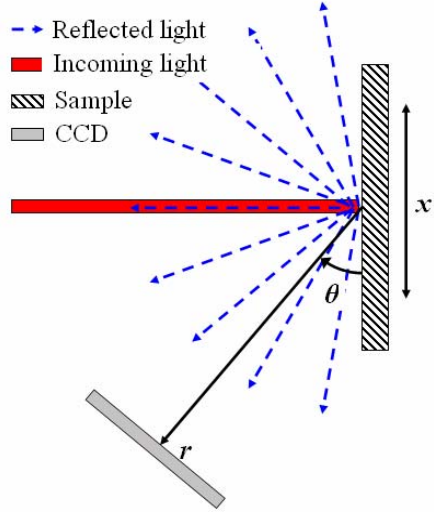

Figure 2.3: The three discussed degrees of freedom of the sample with respect to the incoming light and the CCD: $x$, $r$ and $\theta$.

### 2.3.1 Displacement in *x*-direction

When the sample is slightly shifted in *x*-direction the illuminated part of the sample is changed but the speckle patterns before and after movement will be correlated. The expect that the correlation is caused by the area $O$ that is enlightened in both cases, as depicted in Figure 2.4 on the next page, because the number of rays which have the same phase in both measurements is proportional to $O$.

The area $O$ will now be calculated as a function of the displacement $x_d$. The spot is parameterized as an ellipsoid:

$$\frac{x^2}{a^2}+\frac{y^2}{b^2}=1, \tag{2.6}$$

where $2 \cdot a$ is the short axis and $2 \cdot b$ the long axis of the spot. If the second ellipsoid is moved a distance $x_d$, as one can see in Figure 2.4, the area $O$ can be calculated by integrating $2 \cdot y(x) \cdot dx$ from $x_d/2$ to $a$ and multiply this by 2:

$$\begin{aligned} O &= 2\int_{x_d/2}^{a} 2y(x)dx, \\ &= 2\int_{x_d/2}^{a} \sqrt{(2b)^2\left(1-\left(\frac{x}{a}\right)^2\right)}dx, \\ &= ab\left(\pi - \sin^{-1}\left(\frac{x_d}{2a}\right) - \frac{1}{2}\left(\frac{x_d}{a}\right)^2\sqrt{4\left(\frac{a}{x_d}\right)^2 - 1}\right). \end{aligned} \tag{2.7}$$

Equation 2.7 shows that the area $O$ remains constant when $ab$ and $x_d/a$ remain the same. When the area $A$ of a beam ($A = \pi ab$) is kept constant the overlapping area between two measurements and thus the correlation will be equal for equal values of $x_d/a$.

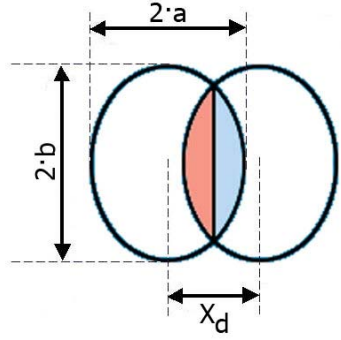

Figure 2.4: Two partially overlapping ellipsoids with short axis *2·a* and long axis *2·b*. The red surface is equal to the blue surface. To calculate the overlap between the two ellipsoids one can integrate only the blue surface and multiply this by 2.

### 2.3.2 Variation in *r*-direction

Changing the distance between the sample and the screen will modify the speckle pattern on the screen. Two effects occur, first the path lengths of the rays change and second the speckles change in size according to formula 2.1.

The effect of the change in the path lengths is considered by estimating the distance $\delta r$ one can move the sample without changing the speckle pattern, see also Figure 2.5. At point $D_1$ the waves from point *P* interfere constructively with the waves from *O* and the waves from *P'* with those from *O'*, etc. When the screen is moved from *r* to *r + δr* one can estimate the change in path length difference $\Delta$ between $PD_1$ and $OD_1$:

$$\Delta = \frac{\partial(OD_1 - PD_1)}{\partial r}\delta r,$$

$$= \frac{\partial}{\partial r}\left(r - \sqrt{\frac{d^2}{2} + r^2}\right)\delta r,$$

$$\approx \frac{\partial}{\partial r}\left(r - r\left(1 + \frac{d^2}{4r^2}\right)\right)\delta r,$$

$$= \frac{d^2}{4r^2}\delta r. \tag{2.8}$$

The bright spot at point $D_2$ will be a dark spot if one sets the change in path length difference $\Delta$ equal to $\lambda/2$:

$$\frac{\lambda}{2} = \Delta$$

$$\approx \frac{d^2}{4r^2}\delta r$$

$$\delta r \approx \frac{2\lambda r^2}{d^2}. \tag{2.9}$$

If one assumes a spot diameter *d* in the order of a few 100 μm, a wavelength $\lambda$ of 632.8 nm and a distance *r* of 10 cm, the distance $\delta r$ is allowed to be in the order of a meter in *r*-direction.

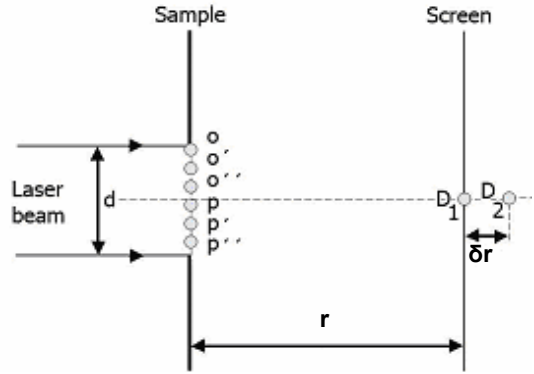

Figure 2.5: A schematic representation for an order of magnitude estimation of the distance one can move the screen such that the speckle patterns are no longer correlated.

The effect of the magnification of speckles is illustrated in Figure 2.6. Due to the large magnification of the pattern the correlation reduces. From (2.1) it is shown that speckles, and therefore the speckle pattern, are magnified when the distance *r* is increased. The effect is negligible when the change in distance is negligible to the distance *r*.

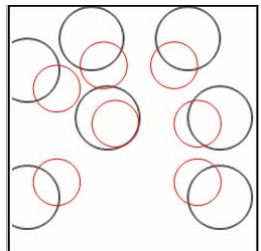

Figure 2.6: The loss of correlation due to the magnification of the speckle pattern when the distance *r* is increased. The speckles are represented by circles and the red speckles represent the image as captured by the CCD at a smaller distance than the black ones.

### 2.3.3 Radial displacement

Placing the sample under a different angle with respect to the recording screen results in a different speckle pattern at the screen. A small misalignment will cause the CCD to record a different part of the speckle pattern. One can find the angle $\theta$ when the CCD, with a width *d*, records a complete new part of the speckle pattern:

$$\theta = \tan^{-1}\left(\frac{d}{r}\right), \qquad (2.10)$$

where *r* is the distance between the sample and the recording screen. At angles larger than $\theta$ there is no correlation expected.

## *2.4 Recognition criterion*

Samples will be recognized by analyzing the correlation function (Appendix D) of two measurements. The correlation functions consist of several peaks which can originate from either a correspondence of the internal structure of the samples or from coincidently coinciding speckles (CCS) in both measurements. From the probability that a peak represents CCS one can judge whether a sample is recognized.

To calculate the probability of a peak representing CCS, Gaussian statistics will be used. The use of Gaussian statistics is justified when events are caused additively and independently, we do not know whether this is the case for correlation functions of speckle patterns. Some further research on this topic is required.

The Gaussian probability distribution is defined as

$$P(x) = \frac{1}{\sigma\sqrt{2\pi}} e^{-\frac{(x-\mu)^2}{2\sigma^2}}, \tag{2.11}$$

where $\mu$ is the average and $\sigma$ is the standard deviation. To determine whether a peak represents CCS the cumulative distribution is used. The probability that there exists a peak $n$ larger than a value $a$ is given by:

$$P(n \geq a) = \int_a^\infty \frac{1}{\sigma\sqrt{2\pi}} e^{-\frac{(x-\mu)^2}{2\sigma^2}} dx. \tag{2.12}$$

When this probability is negligible one can state that the sample is recognized. The probability of finding a peak larger than $a$ depends on the total number of peaks. Assuming that all 512*512 pixels are independent and $N$ correlations are done, the probability of finding a peak larger than $a$ is:

$$P(n \geq a) \cdot N \cdot 512 \cdot 512. \tag{2.13}$$

We define the number of correlations that can be performed before finding a second peak caused by CCS as the uniqueness $U$:

$$U = \frac{1}{512 \cdot 512 \cdot P(n \geq a)}. \tag{2.14}$$

In our research an attempt to distinguish 10 samples will done, this yields 100 correlations. A ratio of 6.5 between the correlation peak and the standard deviation of the correlation is used, corresponding to a uniqueness of $10^5$. Because 100 correlations will be made, there is a probability of 99.9 % that one recognizes the sample properly.

Although speckles are build up out of several pixels, which makes some pixels correlated, this analysis assumes the pixels to be independent. The correlation between the pixels can be taken into account by stating that the number of speckles is the number of independent channels. This approach does not take the position of the speckles into account, which is measurable with an accuracy of the pixel size. Omitting this position effect would overestimate the uniqueness factor whereas our approach underestimates it. How to deal with this problem requires some further research. A simple solution could be to match the speckle size with the pixel size. This solution might lead to experimental problems because the speckles vary in size.

Besides that the pixels are correlated due to the fact that they represent a speckle together, the pixels are always correlated because the CCD is not ideal.

# 3 Experimental apparatus

Because both a line and a spot will be used in the experiments, the setup (see Figure 3) is able to switch easily between illumination by a line and a spot. The light goes either along the path depicted by the dotted line or along the path depicted by the dashed line. The first results in an circular spot and the second in a elliptical spot. One can switch from one configuration to another by turning both flipping mirrors.

The laser light is scattered by the sample and collected by the CCD[5]. The CCD is mounted on a rotational arm (accuracy $1 \cdot 10^{-2}$°). The distance $r$ between the CCD and the sample can be varied (accuracy 0.5mm). The sample is held in a standard clip-type filter holder. This sample holder is connected to the translational stage via screws which allows one to remount the holder within an error of 0.5 mm and an error of 0.5° in the angle under which the light is reflected. Once the sample holder is mounted on the translational stage it can moved along the dotted line with 1 μm accuracy.

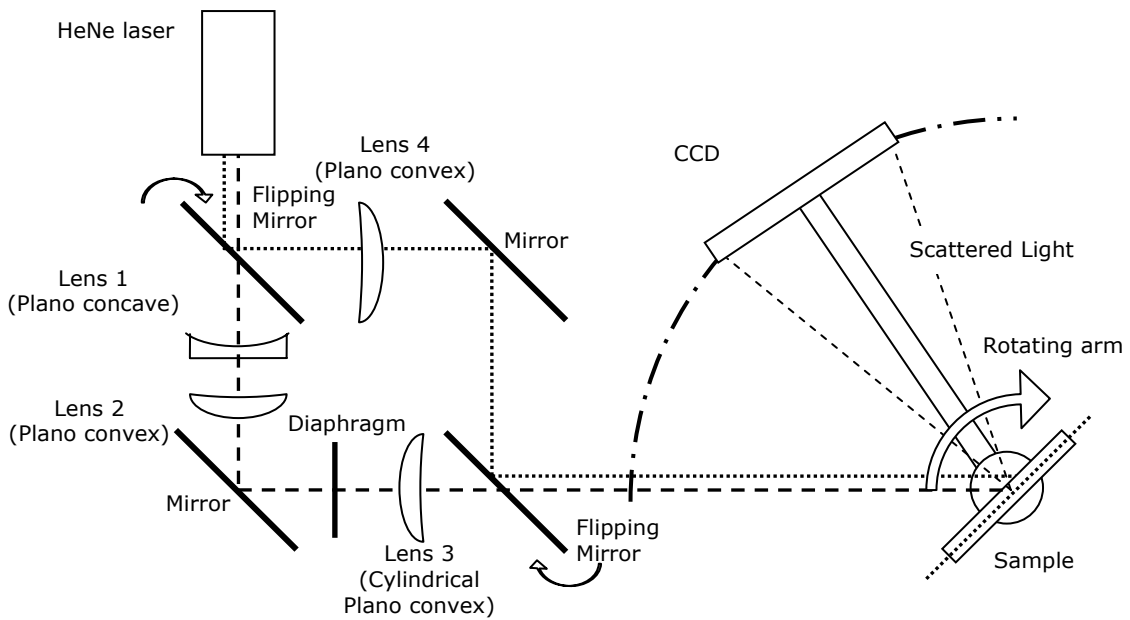

Figure 3.1: A schematic representation of the setup. Lens 1 and lens 2 magnify the beam. Lens 3 focuses the beam to a line on the sample. Lens 4 is used to create a circular spot by focusing the incoming beam onto the sample. The CCD is mounted on a rotational arm allowing measurements of the speckle pattern under different angles and distances.

In the setup a plano convex lens (lens 4) with a focal distance of 200 mm was used to focus the HeNe laser[6] beam to a round spot of 365 ± 5 μm diameter. As shown in Appendix B this is 3.35 ± 0.08 times the diffraction limit.

A spot of 1.93 ± 0.01mm was created by collimating the light with two lenses (lens 1 and lens 2) and using a diaphragm. This spot was focused with a cylindrical lens (lens 3) with a focal length of 60 mm to an elliptical spot of 69 ± 1 μm short axis. The length of the line is 1.65 ± 0.02 times the diffraction limit and the width is 3.35 ± 0.08 times the diffraction limit (see appendix C). Especially the that fact the large side of the line meets the diffraction limit better was expected. One has to note that the diaphragm used to create a line introduces and extra loss in intensity and changes the beam profile. The area of both the elliptical beam and spot are $(1.07 \pm 0.03) \cdot 10^5$ μm², the importance of this has been stressed in 2.3.1.

We used plain white paper samples which were available in the lab. Unfortunately no specifications about these samples were available. One of the samples was measured to be 42.65 ± 0.05 mm by 99.70 ± 0.05 mm and 0.40 ± 0.05 mm thick, its weight was 1.1982 ± 0.0005 g. Thus the paper was 281.8 ± 0.6 g/m². All the samples were made of the same kind of paper.

# 4 Results

In this chapter the final results will be presented in some plots. In the first paragraph the effect of a displacement of the sample in *x*-direction, the sensitivity to a variation in the *r*-direction and measurements on the angular dependence will be presented and discussed. These measurements were done for both the line and the spot.

The recommendations which followed from these measurements are described in paragraph 4.2. Some parameters of the setup were adjusted for the recognition attempt. The sensitivity to a displacement in *x*-direction was measured in this new configuration.

For the recognition attempt the new configuration was used, the results of this attempt are discussed in paragraph 4.3. In this paragraph it is also shown that samples can be recognized with high uniqueness factors, enabling industrial applications.

## *4.1  The measurements on the line and spot*

In this section the sensitivity of the speckle pattern to displacements in *x*, *r* and $\theta$ direction will be discussed and compared to the analyses in chapter 2.

### 4.1.1 Displacement in *x*-direction

In Figure 4.1 the results of the displacement measurement of the 365 ± 5 μm spot are shown. All measurements of the series in the range of 0 μm to 280 μm (20 μm steps) were correlated to the measurements at a displacement of 0 μm, 40 μm, 120 μm, 240 μm and 280 μm of the same series. This resulted in two dimensional correlation functions $C(x,y)$ of which both the peak value and the standard deviation were calculated. Of all correlation functions there are five autocorrelation functions, the 'peak value to standard deviation'-ratios of these functions are plotted at zero displacement.

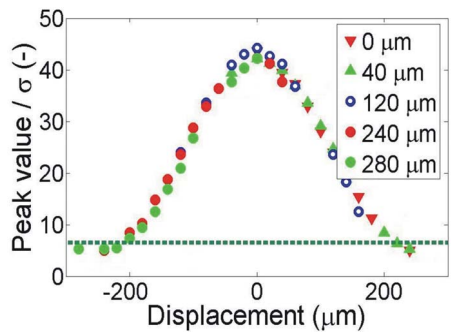 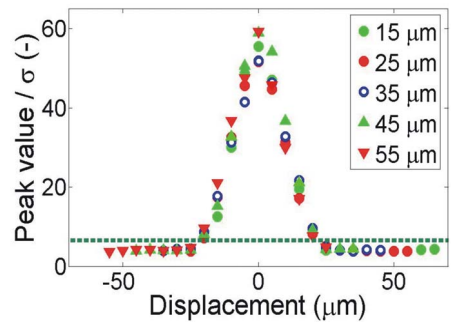

Figure 4.1: The sensitivity of the speckle pattern to a variation in the *x*-direction. The speckle is caused by a circular spot of 365 μm. In the legend one can see to what measurement the shown measurements are correlated.

Figure 4.2: The sensitivity of the speckle pattern to a variation in the *x*-direction. The speckle is caused by a line. In the legend one can see to which measurement the shown measurements are correlated.

Since a 'peak value to standard deviation'-ratio of 6.5 is a good measure for recognition, the sample can be displaced over a distance of minimum 200 ± 10 μm if one uses the spot.

For the line the approach was similar, the results are plotted in Figure 4.2. The measurement series were from 0 μm to 80 μm (5 μm steps) and series were correlated to the measurements at 15, 25, 35, 45, and 55 μm.

Figure 4.2 shows that the line is more sensitive to changes in *x*-direction, even when it is considered relative to the spot's width. The sample can be moved 54 ± 1 % of the spots diameter, whereas the sample can only be moved 25 ± 5 % relative to the lines width. This contradicts the analysis in paragraph 2.3.1 for unknown reasons.

These measurements indicate that when a spot is larger the maximum allowed displacement in *x*-direction, when measured relative to the spot size, is larger too. Since our system can easily produce a 2.62 ± 0.02 mm spot it is recommended to use this spot for the recognition attempt.

### 4.1.2 Variation CCD-sample distance

The influence of a variation of the distance between the CCD and the sample (*r*) was subjected to research at distances of 5.5, 17.6 and 36.7 cm between sample and CCD. We varied the distance *r* with steps of 1 mm. Because the sample can be replaced with 0.5 mm accuracy the variations in the speckle pattern due to a variation in distance can be neglected when the system is insensitive to 1 mm variations.

In Figure 4.3 and 4.4 the 'peak value to standard deviation'-ratio is plotted as a function of the displacement in *r*-direction. Three series were performed, namely at 5.4 cm, 17.5 cm and 32.6 cm. For the measurements shown in figure 4.3 a spot was used and for the measurements in Figure 4.4 a line.

The results in Figure 4.3 show that the system is indeed less sensitive to changes in *r*-direction when the distance between CCD and sample is large. In case of the spot the 'peak value to standard deviation'-ratio remains constant and is sufficient for recognition at a distance of 17.5 cm.

Remarkably, the 'peak value to standard deviation'-ratio is very low when the sample is placed at 17.5 cm and 32.6 cm. The ratio is low due to the broad correlation peak which contributes significantly to the standard deviation of the correlation. The wide correlation peak is caused by the large speckles.

To reduce the speckles in size either the spot has to be increased or the speckles have to be focused. In our current setup the spot size can be increased by a factor of 7.2 to 2.62 ± 0.02 mm. If this spot is used at 30 cm, this will be comparable to measuring at 4.2 cm but with the advantage of lower sensitivity to changes in r-direction.

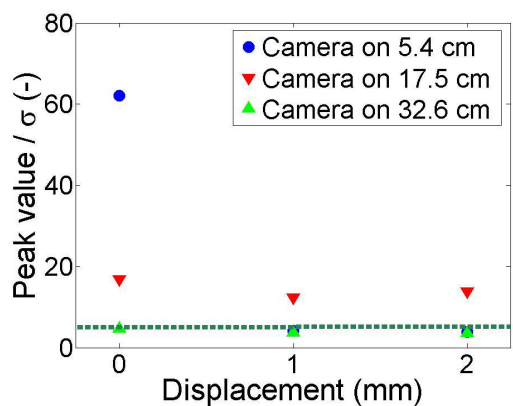
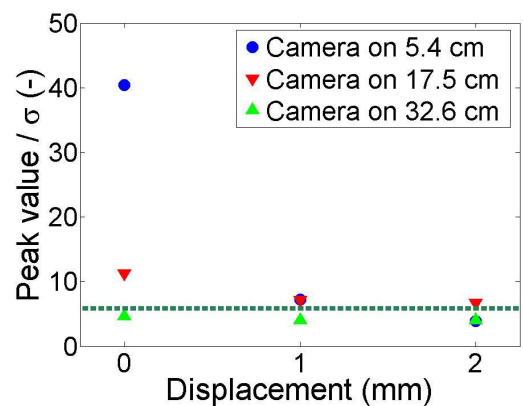

Figure 4.3: In this plot the distance dependence of the spot is shown. At larger distances the 'peak value to standard deviation'-ratio is low, due to the large speckles. At 17.5 cm the correlation remains constant and sufficient for recognition. At 32.6 cm the correlation is too low due to the large speckles

Figure 4.4: In this plot the distance dependence is of the line is shown. At larger distances the 'peak value to standard deviation'-ratio is low, due to the large speckles.

### 4.1.3 Dependence CCD-Sample angle

The angular dependence was measured by varying the angle $\theta$ (see figure 2.3) of the CCD with respect to the sample. A distance of 10.0 ± 0.1 cm was used and the CCD was moved with 0.5° steps. All measurements of this series have been correlated to the measurement at 0° and the result is shown in Figure 4.5. According to our criterion there was no correlation at 3.3 ± 0.3° which corresponds to a movement of 5.8 ± 0.5 mm. It was expected that the correlation would be lost at 6.9 mm, the width of the CCD. We expect there is a correlation between 5.8 ± 0.5 mm and 6.9 mm, but this correlation is probably so small that it can not be distinguished from peaks caused by CCS.

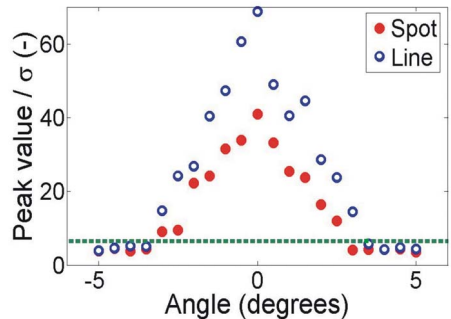

Figure 4.5: In this plot the angular dependence can be seen. As expected the plot is linear. At an angle of 3.3 ± 0.3 there is no significant correlation left.

Furthermore the plot shows that with a line we measure a better peak to standard deviation ratio, the difference is 30 times the standard deviation. This phenomenon can be seen in Figures 4.1 and 4.2 as well, where the difference between the line and the spot is 15 times the standard deviation.

This phenomenon might be explained by the loss in intensity at the diaphragm as discussed in paragraph 2.1, but it requires some further research.

### *4.2 Changes in the setup for successful recognition*

In the former paragraphs two recommendations were done for the recognition measurement, first to increase the distance $r$ to 30 cm and second to use a larger spot. Before the recognition attempt, it was checked whether the larger distance does not change the sensitivity to displacements in the $x$-direction. Measurements on the large spot of 2.62 ± 0.02 mm at a distance of 30 cm were performed. The minimum displacement which would obtain sufficient correlation is 0.9 ± 0.1 mm.

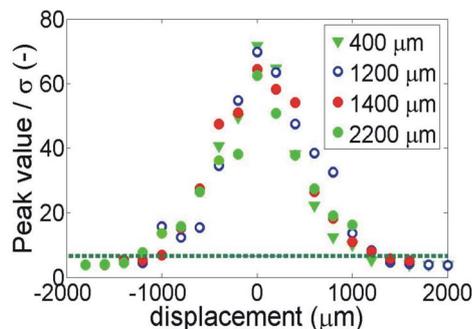

Figure: 4.6: The sensitivity of the speckle pattern to a variation in the position. The speckle is caused by a spot of 2.62 ± 0.02 mm and measured at a distance of 30 cm.

Remarkably, the possible movement relative to the spot size is less for the larger spot. namely 35 ± 4 % of the spot's width instead of 54 ± 1 % . From Figures 4.1 and 4.2 it was expected that the percentage would increase when the spot size was increased. Since both the spot size and the distance *r* were increased we conclude the distance influences the sensitivity in *x*-direction.

## *4.3 Results of the recognition attempt*

With the adjusted setup a recognition attempt was performed. Ten samples were placed in a holder and their speckle patterns were measured. The samples were removed from the translational stage together with the holder. The samples were stored in the holder for thirty minutes in normal laboratory conditions

The first measurement of sample *X* was correlated with all measurements of the second series. The correlation with the highest 'peak value to standard deviation'-ratio was selected. When this was indeed the correlation with the second measurement of sample *X*, and the 'peak value to standard deviation'-ratio was larger than 6.5, the sample is considered recognized. This procedure was done for all measurements of the first series. In Figure 4.7 it is shown that 5 samples have been recognized.

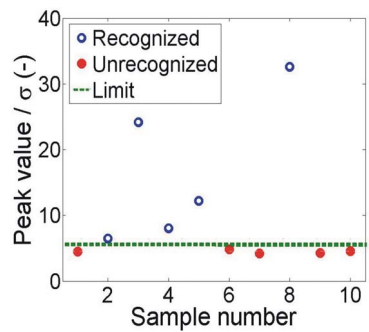

Figure 4.7: The recognition of five out of ten samples. This plot shows the potential of our setup, since very high uniqueness factor have been obtained.

It is interesting to see that the 'peak value to standard deviation'-ratios differ very significantly. Most ratios will not be sufficient for recognizing large numbers of samples. Buchanan *et al.* used a criterion which corresponds to a ratio of approximately 20 for their experiment with 500 samples. All samples which could not be recognized have a peak of approximately 5 times the standard deviation, as expected.

Sample 8 is recognized with a 'peak value to standard deviation'-ratio of 32. The value of 32 indicates that the measurements can be done more precise since Figure 4.6 shows that the maximum possible ratio is 60 and that a ratio of 30 corresponds to a misplacement of 0.5 mm. The fact that the maximum ratio is reduced strongly indicates that the sample holder needs to be improved.

Especially the angle at which the sample is placed was quite critical because the distance between sample and CCD was large. The accuracy in *x*-direction for placing the sample holder was probably problematic too; the accuracy of replacing the sample holder was estimated at 0.5mm. As shown in figure 4.3, the correlation remains constant over a distance *r* for more than 2 mm and therefore variation in *r*-direction is not likely to cause problems.

# 5   Conclusion

In this report the reproducibility of a speckle pattern caused by a line and a spot have been compared. The use of a line always results in a better 'peak value to standard deviation'-ratio. Recognition of 5 out of 10 samples was achieved by using a large spot and a CCD. In our opinion this setup has the potential of distinguishing large numbers of samples.

It was shown that a line is more sensitive to changes in $x$-direction. The sample can be moved 54 ± 1 % of the spots diameter, whereas the sample can be moved only 25 ± 5 % relative to the lines width. This result contradicts the expectation that the sensitivity to a relative movement is independent of beam shape.

The measurements on the distance between the CCD and the sample have shown that it is preferential to work at a large distance, under the condition that the speckles are small enough.

The behavior of the angular dependence appears to be the same in the case of the line and the spot. We found that only the area of the CCD is limiting. At an angle of 3.3 ± 0.3º there is no significant correlation left. This is the point where the chip is moved 5.8 ± 0.5 mm of the chip's width of 6.9mm. We expected a correlation between 5.8 ± 0.5 mm and 6.9 mm, but this correlation is probably so small that it can not be distinguished from peaks caused by CCS.

An attempt to distinguish samples resulted in the recognition of five out of ten paper samples. These results were obtained using a 2.62 ± 0.02 mm circular spot and a CCD. The sample was positioned at 30 cm with respect to the CCD. We suspect that five samples were not recognized because the required accuracy of replacing the samples was not met. The setup has a large potential of recognizing a large amount of samples since recognition with a 'peak value to standard deviation'-ratio of 32 was obtained after thirty minutes and replacing the sample holder. This measurement has not been optimal since it has been shown that a 'peak value to standard deviation'-ratio of 60 can be obtained.

# 6 Recommendations

The setup will yield better results when a sample holder with hardly any angular dependence is used. When the CCD is integrated in the same construction as the sample holder, the angular dependence is strongly reduced. By sliding the paper into the holder and pressing it against a wall, the variation in $x$-direction can be minimized too.

A larger experiment should be performed. When 500 samples are taken a uniqueness factor of $10^{83}$ will be required. This experiment requires the significant improvements of the sample holder mentioned above.

A distance dependence in the sensitivity to displacements in $x$-direction was found. When the 365 ± 5 μm spot was used at 10.0 cm the sample could be displaced 54 ± 1 % of the spot's width. In case of the line this was less, namely 25 ± 5 %. When the spot was increased this percentage was expected to increase too, but it decreased to 35 ± 4 %. Because the distance was increased to 30 cm together with increasing the spot size, a distance dependence is expected. The distance dependence can be explained by reasoning that the CCD does not "see" the borders of the spot anymore when it is very close to the sample. The simple case of an infinite spot illustrates the effect of the border well; the speckle pattern moves as much as the displacement of the sample. In the case of the infinite spot, the correlation becomes proportional to the area of the speckle pattern measured by the CCD before and after movement of the sample.

The fourth recommendation concerns the optimization of the distance between sample and CCD. Increasing the distance is advantageous for variations in $r$-direction, but the angular sensitivity and sensitivity in $x$-direction are smaller when decreasing the distance, therefore an optimum distance can be found.

The last recommendation concerns the fact that the correlations of the line are always stronger than those of the spot. This effect might be caused by the diaphragm; this hypothesis can be verified with a simple experiment.

# Appendix A: Order-of-Magnitude estimation of a speckle

Using a few simple arguments one can make a quick order-of-magnitude estimation for the size of a speckle [4]. For this one needs to suppose that a bright speckle is present at some point $P$ at the screen (see Figure A.1). At this point the light coming from the sample interferes mainly constructively. To simplify the calculation one can assume that, on average, point $D_1$ interferes constructively with $O$, $P$ with $O'$ etc. For an estimation of the speckle size one needs to find a new point, $D_2$, where $P$ and $O'$, $P$ and $O$, etc. interfere destructively.

Taking for example points $P_1$ and $P_2$, one needs to find the change $\delta x$ such that the corresponding change in the path length difference $OD_1 - PD_1$, $\delta(OD_1 - PD_1)$ is equal to $\lambda/2$. Since $OD_1$ is equal to $(x^2+L^2)^{1/2}$ and $PD_1$ is $(((d/2)-x)^2 + L^2)^{1/2}$, one obtains, together with the assumption that $d<<L$,

$$\delta(OD_1 - PD_1) =$$
$$\delta\left(\sqrt{x^2+L^2} - \sqrt{\left(\frac{d}{2}-x\right)^2 + L^2}\right) \cong$$
$$\delta\left(L\left(1+\frac{x^2}{2L^2}\right) - L\left(1+\frac{d^2}{4L^2} - \frac{dx}{L^2} + \frac{x^2}{L^2}\right)\right) =$$
$$\delta\left(\frac{d^2}{4L} - \frac{dx}{L}\right) =$$
$$\frac{d}{2L}\delta x. \tag{A.1}$$

Setting this equal to $\lambda/2$ leads to an expression for $\delta x$ and from there one can write the following expression for the speckle size $\sigma$, which is two times $\delta x$:

$$\sigma \cong 2\delta x \cong \frac{2\lambda L}{d}. \tag{A.2}$$

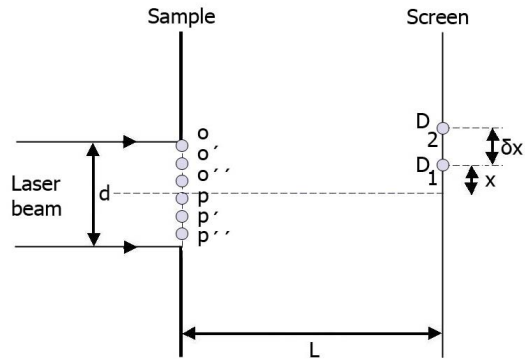

Figure A.1: A schematic representation for an order of magnitude estimation of the speckle size. When the path length difference from two edges of the enlightened sample to the speckle is half a wavelength The speckle will change from bright to dark or vice versa.

## Appendix B: Diffraction limited size of the spot

In this appendix the calculations on the spot size of our setup will be presented.

The circular spot was produced by focusing the laser light with a 200 mm lens after it had diverged over a distance of 1.05 m. At the lens the waist is given by the following formula;

$$w(1,05) = w_{0,l}\left[1+\left(\frac{\lambda 1,05}{\pi w_{0,l}^2}\right)^2\right]^{1/2}, \tag{B.1}$$

where $w_{0,l}$ is the waist of the laser, 0.315 mm and λ is 632.8 nm. This beam is then focused to the waist $w_{0,f}$ by a 200 mm lens:

$$w(1,05) = w_{0,f}\left[1+\left(\frac{\lambda f}{\pi w_{0,f}^2}\right)^2\right]^{1/2}, \tag{B.2}$$

Using this equation in combination with B.1 the diffraction limited spot size can be calculated, which is 54.5 ±0.5 μm.

# Appendix C: Diffraction limited size of the line

In this appendix the calculations on the line in our setup will be presented.

The laser light is first diverged in the first part of the setup, over a distance of 24 cm:

$$w(0.24) = w_{0,l}\left[1+\left(\frac{\lambda 0.24}{\pi w_{0,l}^2}\right)^2\right]^{1/2}. \tag{C.1}$$

Then the light is expanded by two lenses. First the beam is further diverged by a negative lens with a -75mm focal distance; it propagates over a distance of 175mm. Then the light is focused by a 250 mm lens. The expansion of the beam can be represented by a 75mm lens $f_{t,1}$ which focuses the light to a waist $w_{0,t}$ and then a 250 mm lens $f_{t,2}$ which focuses it at 325mm. The waist $w_{0,t}$ can be extracted from:

$$w(0.24) = w_{0,t}\left[1+\left(\frac{\lambda f_{t,1}}{\pi w_{0,t}^2}\right)^2\right]^{1/2}. \tag{C.2}$$

The size of the spot at the second lens at 0.415 m is then given by:

$$w(0.415) = w_{0,t}\left[1+\left(\frac{\lambda f_{t,2}}{\pi w_{0,t}^2}\right)^2\right]^{1/2}. \tag{C.3}$$

Since the beam is collimated now we may assume this value to be the waist of the collimated beam. This beam diverges over a distance of 83.5 cm, which gives the length of the ellipse.

$$w_{ellipse} = w(0.415)\left[1+\left(\frac{\lambda 0.835}{\pi w(0.415)^2}\right)^2\right]^{1/2} \tag{C.4}$$

When performing these calculations one obtains a length of 584±4 μm. All components can be placed with 1 cm accuracy. By filling in al the distances plus or minus a centimetre one can estimate the error.

To create a line from this round spot the light is focused by a 6cm cylindrical lens after 77,5 cm on 1,19m from the laser, where the beam is:

$$w(1.19) = w(0.415)\left[1+\left(\frac{\lambda 0.775}{\pi w(0.415)^2}\right)^2\right]^{1/2}. \tag{C.5}$$

The thickness $t_{ellipse}$ of the ellipse is given by:

$$w(1.19) = t_{ellipse}\left[1+\left(\frac{\lambda f_c}{\pi t_{ellipse}^2}\right)^2\right]^{1/2} \tag{C.6}$$

These calculations give a value of 10.3 ±0.1 μm.

## Appendix D: Correlations

Cross correlation is a standard method of estimating the degree to which two signals are similar. When one wants to have a quantitative measure of shared properties of two random signals one can use such a correlation function. The cross correlation may be defined as:

$$c_{fh}(\tau) = \int_{-\infty}^{\infty} f^*(t) \cdot h(t+\tau) dt \tag{D.1}$$

Using the theory of Fourier transforms this can be written as:

$$\int_{-\infty}^{+\infty} f^*(t) \cdot h(t+\tau) dt =$$

$$\int_{-\infty}^{+\infty} \left[ \frac{1}{2\pi} \int_{-\infty}^{+\infty} F^*(\omega) \cdot e^{+i\omega t} d\omega \right] \cdot h(t+\tau) dt =$$

$$\frac{1}{2\pi} \int_{-\infty}^{+\infty} F^*(\omega) \left[ \int_{-\infty}^{+\infty} h(t+\tau) e^{+i\omega t} dt \right] d\omega =$$

$$\frac{1}{2\pi} \int_{-\infty}^{+\infty} F^*(\omega) H\{h(t+\tau)\} d\omega =$$

$$\frac{1}{2\pi} \int_{-\infty}^{+\infty} F^*(\omega) H(\omega) e^{-i\omega t} d\omega \tag{D.2}$$

So in order to calculate a cross correlation one can first Fourier transform both signals, take the conjugate of one, multiply them and then do an inverse Fourier transform.